\def\P1{{\bf P}^1}
\def\Pt1{${\bf P}^1$}

\def\S{{\Sigma}}

\input harvmac
\Title{\vbox{\hbox{hep-th/9703194}}}
{\vbox{\hbox{\centerline{F Theory on $K3\times K3$ and}}
\vskip .3in
\hbox {\centerline{ Instantons on 7-branes}}}}
\vskip .3in
\centerline{\sl Michael Bershadsky}
\vskip .2in
\centerline{\it Lyman Laboratory of Physics, Harvard
University}
\centerline{\it Cambridge, MA 02138, USA}
\vskip .3in
\centerline{\sl Vladimir Sadov}
\vskip .2in
\centerline{\it  Institute for
Advanced Study}
\centerline{\it Princeton, NJ 08540, USA}

\vskip .3in
\centerline{ABSTRACT}
\vskip .2in
We discuss $N=2$ supersymmetric compactifications to four dimensions from
the point of view of F-theory and heterotic theory.
In a relatively
simple
setup, we illustrate the spectral theory for vector bundles on $K3\times
T^2$ and discuss the heterotic -- F-theory map. The moduli space of
instantons on the 7-brane wrapped around $K3$ is discussed from the point
of view of Higgs mechanism in the effective four-dimensional $N=2$
theory.
This allows us to elaborate on the
transitions between various branches of the  moduli space of the
heterotic/F-theory. We also describe  the F-theory compactification
on smooth $K3
\times K3$ without the 3-branes in which the anomaly is cancelled
by the gauge
fields.
\Date{March, 1997}

\newsec{Introduction}

F-theory has proved  to be a very effective  tool in studying
string theory in
$D \geq 5$
dimensions \ref\vafa{C. Vafa, Nucl.Phys. {\bf B469} (1996) 403}
\ref\mv{D. Morrison and C. Vafa, Nucl.Phys. {\bf B473} (1996) 74-92}
\ref\mvv{D. Morrison and C. Vafa, Nucl.Phys. {\bf B476} (1996) 437}.
Partially this can be attributed to relatively simple dynamics of
higher-dimensional gauge theories, which can be captured by the
geometry of
elliptic fibrations. Compactifications down to four
dimensions are different in this respect. Various
perturbative and nonperturbative phenomena occur in the effective
field theory,
some of them do not have any straightforward interpretation in terms
of elliptic
fibrations. The first question that arises is simply whether there
is a moduli
space, or all flat directions are lifted due to the superpotential.
At present,
we do not know how to answer this question completely, although the
mechanism
for generating the superpotential is known and it the
superpotential is shown
to be generated in some examples
\ref\per{E. Witten, Nucl.Phys. {\bf B474} (1996) 343} \ref\pan{M.
Bershadsky,
A. Johansen, T. Pantev ,V. Sadov and C. Vafa, {\it F-theory, Geometric
Engineering and N=1 Dualities} hep-th/9612052}.

In this paper we discuss the F-theory compactification on $K3 \times K3$
which possesses  $N=2$ supersymmetry  in four dimensions. This
allows to avoid
the question about the superpotential and makes the discussion of
the moduli
space much easier.
Similarly to a generic F-theory compactification on the Calabi-Yau
fourfold,
the compactification on $K3\times K3$ has a three-brane anomaly of
$-24$ which
can be cancelled by either $24$   3-branes \ref\vws{S. Sethi, C.
Vafa and E.
Witten, Nucl.Phys. {\bf B480} (1996) 213.} or
by the nontrivial background gauge fields on the 7-branes. Thus
there are two
kinds of  moduli, the ones associated with 3-branes and the others
associated
with 7-branes. The 3-brane in the bulk (away from 7-branes) carries
an $N=4$
vector multiplet which consists of $N=2$ vector multiplet and a
hypermultiplet
in the adjoint. The corresponding gauge theory is in the Coulomb
phase with the
complex scalar of the vector multiplet measuring the position of
the 3-brane
with respect to 7-branes. The 7-brane carries an $N=2$ vector
multiplet as well
as hypermultiplets in various representations. The background gauge
bundle on
the compact part of the 7-brane worldvolume breaks the gauge
symmetry down from
the maximal one dictated by the singularity of elliptic fibration.
In terms of
the effective field theory in four dimensions this symmetry
breaking can be
interpreted as Higgs mechanism in the maximal gauge theory, which
leads to
identification of the Higgs branch of moduli  space
with the moduli space of instantons. In fact, we will see that this
identification is incomplete in the sense that the moduli space of
instantons
carries a bit more information about the string theory compactification.

As a 3-brane approaches the 7-brane, the other kind of
hypermultiplets become
massless. They come from strings connecting 3-brane to the  7-brane. The
expectation values of these hypermultiplets become a part of
coordinates of
the Higgs moduli space.  Since the 3-brane on top of the 7-brane
with these
expectation values turned on is indistinguishable from an instanton
in the
background gauge bundle on the 7-brane, we refer to this process as
the 3-brane
-- instanton transition \ref\smal{E. Witten, Nucl. Phys. {\bf B460}
(1996) 541}
\ref\dog{M. Douglas, {\it Gauge Fields and D- Branes},
hep-th/9604198}. This
transition does not change the total value of the 3-brane anomaly.

In general in $N=2$ supersymmetric gauge theories the
(hyperk\"ahler) Higgs
branch does not receive quantum corrections while the Coulomb
branch does. Thus
it is important to understand to what extent the above geometric
description
accounts for these corrections. As it will become clear, the
geometric Coulomb
moduli space of 3-branes is not
corrected
while the geometric moduli space of 7-branes (the complex structure
of elliptic
fibration) does get quantum corrections. In particular, the geometric
singularity of elliptic fiber on the 7-brane predicts certain gauge
symmetry.
When it is unbroken by the background gauge bundle, i.~e.~when the
gauge theory
is in the Coulomb phase, this symmetry is broken by the strong coupling
infrared effects which generate  the mass gap. The value of the
mass gap is
related to the K\"ahler rather than complex structure of $K3\times
K3$ so one
may speculate about the existence of some more general geometric
theory that
would combine both to reproduce the quantum corrections.

There is also another interesting possibility. Consider two 7-branes both
equipped with the nontrivial gauge bundles approaching each other
so that by
the end of the day there are two 7-branes on top of each other. It
turns out
that in many cases this
configuration
does not produce a  gauge group in four dimensions because of the
gauge bundle inside the 7-branes.
Naively it looks like that by adjusting the relative positions of
the 7-branes
(parameters of
the Coulomb branch) we end up far in the middle of the Higgs branch.
We explain this phenomenon in the Section 4 noticing that the same
space (the
moduli space of instantons on $K3$) may be viewed as Higgs branch
for several
different Coulomb branches. More precisely, looking from the
Coulomb branch of
the gauge theory with maximal gauge group, the Higgs branch is an open
{\it patch} of the instanton moduli space in the vicinity of small
instantons.
The instanton moduli space is  a way the string theory
compactifies the Higgs branch; in doing this some points at
infinity may be
added.
The Coulomb branches of other gauge theories with smaller gauge
groups touch
the  compactified Higgs branch
exactly at these points.
In the example considered in this paper the $SU(2)$ Higgs
branch with the even number of hypermultiplets touches the $U(1)$ Coulomb
branch.

The inverse process is also very interesting. At the  special
points in the
Higgs branch  where the gauge bundle is reducible one can deform the
singularity $I_{G_1 \times G_2} \longrightarrow I_{G_1} \times
I_{G_2}$.  As
the result of this process one obtains two 7-branes equipped with
the gauge
bundles with {\it non zero} first Chern class.
The typical example of this situation would be
the deformation of $I_2$ singularity into two $I_1$ singularities.
Therefore,
starting with the compactification with 24 7-branes and 24 3-branes we
first can go to the enhanced symmetry locus, in which some of the
7-branes
are on top of each other, then replace 3-branes by nonabelian instantons
and  further deform a theory to a locus in which the gauge bundle is
reducible.  Repeating this process one can reach the phase with
24 7-branes equipped with nontrivial line bundles and  with no
3-branes.  In this phase, the fourfold $K3\times K3$ is {\it smooth}.

Another question that should be understood is the heterotic/
F-theory duality
in four dimensions. Some progress in understanding the general aspects of
duality was recently achieved \ref\morg{R. Friedman, J. Morgan and
E. Witten,
{\it Vector Bundles And F Theory}, hep-th/9701162} \ref\bp{M.
Bershadsky, A.
Johansen, T. Pantev and V. Sadov, {\it On Four-Dimensional
Compactifications of
F-Theory}, hep-th/9701165}. It turns out that the complex structure
of the
elliptic fibration on the F-theory side codes both the complex
structure of the
heterotic Calabi-Yau threefold and the a part of the information
about the
heterotic vector bundle known as {\it spectral cover}. The gauge
bundles inside
7-branes are mapped into the other piece of data specifying the heterotic
bundle known as {\it spectral bundle}. Together, the spectral cover
and the
spectral bundle fix the heterotic bundle completely. Also, the
F-theory/heterotic duality maps 3-branes into heterotic 5-branes.
All this  turns out to be strikingly  simple for the $K3\times K3$
compactification.  The dual heterotic theory is compactified on $K3
\times
T^2$. It can be viewed in two different ways. We can
either first
compactify down to $6$ dimensions on $K3$ and then further on $T^2$, or
compactify first down to $8$ dimensions on $T^2$ and then further
on $K3$. The
second point of view appears to be very fruitful. In this case the
spectral
theory is very simple and the spectral surface is just a collection
of $K3$
surfaces.
On the F-theory side the set of 7-branes is also  a collection of
$K3$s. This
observation  also simplifies the discussion of the map between the
gauge bundle
on
different components of the spectral surface and the gauge bundle
inside the
7-branes.

There is another duality which maps  F-theory compactification on
$K3\times K3$
to Type IIA compactification on the certain elliptic Calabi-Yau
threefolds. The
base of fibration is a (generalized) Del Pezzo surface which is
$\P1\times \P1$
blown up in $N$ points. As usual in Type IIA, the Higgs branch is
parameterized
by the complex structures and the Coulomb branch is parameterized by the
K\"ahler structures of the threefold. It turns out that under the
duality the
Coulomb branch of vector multiplets on 7-branes is mapped to the K\"ahler
moduli of singular fibers, while the Coulomb branch of vector
multiplets on
3-branes is mapped to the K\"ahler moduli moduli of $N$ blow ups of
the base.
Thus the number $N$ corresponds in F-theory to the number of
3-branes in the
Coulomb phase, i.~e.~sitting in the bulk away from  7-branes.
Preserving the
Calabi-Yau condition, one can blow up as many as $24$ points on
$\P1\times \P1$
if these points lie on two parallel lines. The resulting Calabi-Yau
has Hodge
numbers $(43,43)$ and two $E_8$ singularities along two rational
curves. It is
dual to F-theory with all $24$ 3-branes in the bulk.

\newsec{Chain of dualities vs. the anomaly.}

We will start by reviewing the chain of dualities associated with
 the  F-theory compactification on $K3\times K3$ \vws. Doing the
compactification
on
one $K3$ first we get a dual of heterotic theory on $T^2$.
Compactification on the second $K3$ gives heterotic theory on $T^2\times
K3$. On the other hand, since heterotic on $K3$ is F-theory on Calabi-Yau
threefold $CY_3$, the theory at hand is dual to F-theory on $T^2\times
CY_3$ which is Type IIA on $CY_3$.

Naively,
there appears to be a puzzle
when one realizes that the F-theory on $K3\times K3$ has an anomalous
3-brane
charge 24. This anomaly can be derived directly from the D-brane picture.
The base
$B_F=\P1 \times K3$, and the discriminant locus consists of 24 D-branes
represented by 24 points on \Pt1 and wrapped on $K3$. Wrapping the
7-brane
around $K3$ produces a 3-brane charge of $-1$, so in total there is
$Q_{(3)}=-24$ to be compensated. To cancel the anomaly one can put 24
3-branes into $K3\times K3$. Thus in the dual heterotic picture one
needs to turn on  24 5-branes wrapped around $T^2$.
(What happens in the Type IIA picture, we will discuss below.)
However, there are heterotic  compactifications on $T^2\times K3$ without
5-branes connected  by transitions to compactifications with  
5-branes. What
these compactifications correspond to in the F-theory description?

In fact, this apparent puzzle can be explained completely  within the
heterotic
description. To obtain a
 compactification on $T^2\times K3$  we can
compactify either first on $K3$ and then on $T^2,$ or first on $T^2$ and
then on $K3$. Following the first path, {\it generically} we choose a
$(12-n,12+n)$
$K3$ compactification with $244$ massless hypermultiplets, no vectors and
one tensor multiplet (for $n=0,1,2$). Compactifying further on $T^2$
we recover the $N=2$ theory with $244$ hypermultiplets and 3 vectors.
Since the $E_8\times E_8$ is  already broken completely in 6 dimensions,
there are no Wilson lines to put on $T^2$. Thus the corresponding
$E_8\times E_8$ heterotic bundle on $T^2\times K3$ is a pullback of the
bundle on $K3$. The heterotic anomaly is cancelled by the second Chern
classes of $E_8$ bundles.

If we compactify on $T^2$ first, {\it generically} we get an $N=1$ theory
with $18$ vectors in eight dimensions. The group $E_8\times E_8$ is
broken
to $U(1)^{16}$ by the non trivial Wilson lines around $T^2$ and the
Kaluza-Klein modes of $T^2$ provide two more $U(1)$'s. Let us try to
compactify this theory on $K3$, choosing the trivial background
$U(1)^{18}$ bundle, so that the heterotic bundle on $T^2\times K3$ is a
pullback from the bundle on $T^2$. Then to compensate for the
second Chern
class of $K3$, one needs to turn on 24 heterotic 5-branes wrapped around
$T^2$. The resulting
theory generically has an unbroken $U(1)^{18}$ from eight dimensions plus
$U(1)^{24}$ from the 5-branes. It is this compactification that  can
be identified  with  F-theory on $K3\times K3$ with $24$ 3-branes \vws.

We may also choose a different background for $U(1)^{18}$, picking $18$
 line bundles $L_i$ on $K3$ such that $\sum c_1(L_i)=0$ and $\sum
c_1^2(L_i)=-48+2N_5$. The heterotic anomaly is cancelled by the gauge
fields and $N_5$ 5-branes. As the above examples show, there seems
to be a
discrete set of combinations of a number of 5-branes and a set of the
Chern classes for the background gauge bundles to cancel the anomaly. In
fact, all these compactifications are connected on the larger
moduli space
of all heterotic compactifications, as we show below.

\newsec{Bundles on $T^2\times K3$}

 Having followed two different paths to get to four dimensions, one ends
up with two rather different types of heterotic bundles on
$T^2\times K3$.
To put these two examples in a general framework we will consider on
$T^2\times K3$ the bundles with the first Chern class equal to zero and
the second Chern class a pullback from $K3$. For simplicity, we will
discuss only the $SU(r)$ bundles.

To describe the vector bundles, we use the spectral theory, which
dramatically simplifies on the product $T^2\times K3$. A vector
bundle $V$
is described by means of two objects: a spectral surface and a collection
of spectral bundles.
The {\it spectral surface} $\S(V)$ of
bundle $V$  consists of several components $\cup_i n_i x_i\times S_i$,
where $S_i$ is a copy of $K3$ and $x_i$ is a point in the dual torus
$\check{T}^2$, $n_i$ is its multiplicity so that $\sum n_i=r$. Each
component $S_i$ carries  a {\it spectral bundle} $M_i$ of rank $n_i$. The
bundle $V$ on $T^2\times K3$ is fixed by this data as follows:
\eqn\spec{
V=\bigoplus_i p_1^*L_{x_i}\otimes p_2^*M_i,
}
 where $L_{x_i}$ are  the line bundles on $T^2$ corresponding to $x_i\in
\check{T}^2$. We denote by
$p_1$ and $p_2$ the projections from the product $T^2\times K3$ to the
factors $T^2$ and $K3$, respectively.

The first Chern class of $V$
is zero
which forces $\sum_i n_ix_i=0$ and $\sum_i c_1(M_i)=0$, but the
individual
$c_1(M_i)$'s may be
non-trivial. The second Chern class $c_2(V)$ is given by
\eqn\sechc{
c_2(V)=\sum_i c_2(M_i)-{1\over 2}c_1^2(M_i).}

When all $x_i=0\in \check{T}^2$, there is only one component of the
spectral surface and the bundle $V=p_2^*M$ is a pullback of the
bundle $M$
on $K3$. Alternatively, when all $x_i$ are different the bundles $M_i$
are  the $U(1)$  bundles which appear in the second
compactification path above.

Let us study the deformations of $V$. They come from the deformations of
the spectral surface $\S(V)$ and the deformations of the collection of
spectral bundles $M_i$. If the component $x_i\times S_i$  of $\S(V)$ has
multiplicity $n_i=1,$ it contributes one-parameter family of
deformations
corresponding to moving $x_i$ on the dual torus $\check{T}^2$. On the
other hand, if $n_i \geq 2$ there are three kinds of deformations. {\it
First},
one can again  move $x_i$ on $\check{T}^2$. {\it Second}, one can deform
the
spectral bundle $M_i$: the dimension of the moduli space (for $U(n_i)$
bundles) is given by
\eqn\dimmod{
{\rm dim}\,{\cal M}_{M_i}
=2n_i\,c_2(M_i)-2(n_i^2-1).
}
 {\it Finally}, the deformations of the third kind  split the multiple
point to
several points with smaller multiplicities. Such deformations are only
possible if the spectral bundle $M_i$ is reducible: $M_i=\oplus_j
M_{ij}$.
Then one can split the component $n_i x_i\times S_i$ to $\cup_j\ {\rm
rk}(M_{ij})\, x_{ij}\, S_{ij}$.

As an example, let us consider the splitting of a multiplicity $2$
component with an $SU(2)$ spectral bundle $M$ on it. The splitting may
occur in the (singular) points of the moduli space ${\cal M}_{SU(2)}$
where $M=L\oplus L^{-1}$. The possible line bundles $L$ are restricted by
the
condition $c_2(M)=-c_1^2(L)$. Note that $c_2(M)$ has to be even since
$c_1^2(L)$ is always even. The multiplicity $2$ component with $M=L\oplus
L^{-1}$ splits into
two multiplicity $1$ components carrying the spectral bundles $L$ and
$L^{-1}$ respectively.

This procedure is a regular way to produce
spectral
bundles with the non-trivial first Chern classes. For instance, one may
start with a pullback of a $(12-n, 12+n)$  $E_8\times E_8$ bundle
on $K3$,
go to the point on the moduli space where this bundle splits, then deform
and end up with 16 line bundles with non-trivial $c_1$'s. The gauge
symmetry is restored to $U(1)^{18}$. This model  describes an alternative
compactification of eight dimensional heterotic theory down to four
dimensions, where the $K3$ anomaly ${\rm tr}\,R^2$ is cancelled by
$\sum_i
c_1^2(L_i)$ of the background $U(1)$ bundles and the 5-branes are absent.

Now one may ask if it is possible to arrive at this compactification
starting with the compactification with the 5-branes. Here we will
outline
the answer leaving the detailed discussion for the next section. Let us
interpret a 5-brane as a point-like gauge  instanton on
$K3$. In a 5-brane -- instanton transition the instanton  acquires finite
size. However, in the vicinity of  the pointlike instantons, the
bundle is irreducible. One has to go a finite distance on the
moduli space
to reach the point where the bundle splits. Therefore, the transition
connecting two heterotic vacua, one with 5-branes  and the
other without 5-branes, goes via the locus of gauge symmetry
enhancement along the moduli space of  instantons on $K3$.

\newsec{Bundles on 7-branes}
\subsec{Heterotic -- F-theory map}

This model admits  a  particularly simple map
between heterotic and F-theory pictures. On the general grounds, we
expect
that
the 5-branes are mapped into the 3-branes, the spectral surface is mapped
into the complex structure of the elliptic fibration and the spectral
bundles are mapped into the gauge bundles on 7-branes. In our
example, all
these maps are straitforward:

The F-theory 3-branes in
$\P1\times
K3$ are mapped to heterotic 5-branes  wrapped on $T^2$ in
$T^2\times K3$. The
map identifies two $K3$ factors and interprets \Pt1\ as a Coulomb
branch of the
effective worldvolume theory of the 5-brane wrapped on $T^2$.

The map between the spectral surface $\S(V)$ and the discriminant
locus follows
from the heterotic/F-theory duality in 8 dimensions. Indeed, $\S(V)$ is
determined by the Wilson lines $V\mid _{T^2}$ which together with
the complex
and K\"ahler structure of $T^2$ determine the heterotic
compactification on
$T^2$. The discriminant locus is on the other hand  determined by
$24$ points on
\Pt1 which describe the dual F-theory compactification on $K3$.

Generically, both the spectral bundles and the bundles on 7-branes
have rank
one. The heterotic $E_8\times E_8$ is broken to $U(1)^{16}$ so  that the
spectral surface consists of $16$ copies of $K3$. There are two other
$U(1)$ vector
multiplets corresponding to the complex and Ka\"hler moduli of $T^2$.
In F-theory, there are $24$ mutually non-local 7-branes with $18$
independent $U(1)$'s.
These  line bundles are mapped into the $18$ line bundles on the
heterotic
side.

 Now
consider the process in which two components of the spectral
surface join
into one component with the multiplicity two: $(x_1 S_1) \cup (x_2 S_2)
\rightarrow 2x S$.
The above map between the  spectral surface and the  discriminant
tells us
that at this very
moment two 7-branes with singularities $I_1$ join into
one 7-brane with singularity $I_2$, so one can identify the
rank two bundle on the component of the spectral surface with the
rank two
 background bundle defined on the 7-brane with singularity $I_2$.
It is easy
to continue
this identification.  The spectral bundles $M_i$ should be mapped
into the
background gauge bundles on   the 7-branes. In particular, the $SU(n)$
gauge symmetry enhancement can be either described in terms of $n$
7-branes coming together or $n$ components of the spectral surface coming
together to form a multiplicity $n$ component. In this situation,
there is
a one-to-one correspondence between the components of the spectral
surface  and the spectral bundles on one hand, and the 7-branes with the
background gauge bundles on the other hand \bp.

 Because of that correspondence, everything that was said above
about the spectral bundles can also be applied to the background gauge
bundles on 7-branes. In particular, a (multiple) 7-brane equipped
with the {\it
irreducible} bundle cannot split into a union of 7-branes with smaller
charges. If the bundle splits as a direct sum of several
sub-bundles $M_i$, the
7-brane can split to several 7-branes, each equipped with its own $M_i$.

\subsec{3-branes and  structure of the moduli space}
Let us discuss the 3-brane -- instanton transition in this context,
stressing the new aspects the compactification brings in. Consider an
$A_{N-1}$ 7-brane with the
background bundle $M$ on it. In four dimensions this gives rise to a pure
$N=2$ supersymmetric $SU(N)$ Yang-Mills theory.  A 3-brane on top of the
7-brane contributes a massless hypermultiplet in the fundamental
(antifundamental) representation of $SU(N)$. Let us take a number
$k>N$ of
such 3-branes, so that there are $k$ hypermultiplets and the gauge group
can be Higgsed completely.
The (baryonic) Higgs branch has the dimension
\eqn\higd{2Nk-2(N^2-1).
}
 As a space, it is a hyperk\"ahler quotient $\cal H$ of the total
space of
matter fields ${\bf C}^{2Nk}$ by the action of the complexified gauge
group $SL(N,{\bf C})$.

On the other hand, a 3-brane on top of the
$SU(N)$ 7-brane can be identified with a pointlike instanton (a
torsionless sheaf, to be more precise). Higgsing may be interpreted along
the lines of the ADHM construction as giving this instanton a
finite size.
Therefore we are led to an identification of the baryonic Higgs branch
with the moduli space of $SU(N)$ instantons on $K3$ with the instanton
number $k$. Comparing the dimensions \dimmod\ and \higd\ one sees
they are
the same. Also, both spaces are hyperk\"ahler.

However, it would be wrong to claim these spaces are identical. Indeed,
the (Gieseker) moduli\foot{There are at least two different
compactifications
of the moduli space of instantons on $K3$. In the string theory
context it is
appropriate to consider the Gieseker compactification, which adds {\it
torsionless sheaves} to vector bundles to get the compact space. The
torsionless sheaves correspond to pointlike instantons considered
as 3-branes.
} space $\cal M$ of $SU(N)$ bundles on $K3$ is a {\it compact}
variety, which cannot be obtained by a naive ADHM-like construction.
Rather, the field-theoretical Higgs branch $\cal H$ lies in $\cal
M$ as an
open neighborhood of the point-like $k$-instantons. String theory
provides a natural compactification for $\cal H$ completing it to $\cal
M$. The compactification may add certain singularities which describe the
physics missed by the naive field theory description. As a concrete
example, let us consider $SU(2)$. For {\it even} $k$, the moduli space
${\cal M}_{SU(2)}$ has conifold singularities at points corresponding to
reducible bundles $L\oplus L^{-1}$, where $L$ are line bundles with
$-c_1^2(L)=k$. From the point of view of four dimensional theory,
singularities at these  points correspond to the fact that the
gauge group
is restored to $U(1)$. At these points, the $U(1)$ Coulomb branch touches
${\cal M}_{SU(2)}$. This Coulomb branch is different from the one
attached
at the origin of $\cal H$: on it the number of 3-branes is smaller by
$-c_1^2(L)$.  It corresponds to splitting of the $SU(2)$  7-brane
into two
$U(1)$ 7-branes, with the background $U(1)$ bundles $L$ and $L^{-1}$
respectively. The complex scalar parameterizing the Coulomb branch
measures the separation between two 7-branes.

It is worthwhile to discuss the effective U(1) theory. At the point
where the
Coulomb branch touches the Higgs branch, there are massless fields coming
from the moduli of the $SU(2)$ instanton bundle on $K3$. If this point
{\it were} smooth on ${\cal M}_{SU(2)}$,  these fields would form the
tangent
space with the dimension  equal to $4k-6$. The tangent space to ${\cal
M}_{SU(2)}$ at a smooth point corresponding to a vector bundle $M$ is
given by the cohomology group $H^1(K3, End(M))$. However, the point
$M=L\oplus L^{-1}$ is not smooth, there is a conifold singularity
there. The
cohomology group
\eqn\zaris{
H^1(End(L\oplus L^{-1}))=H^1(L^2)\oplus H^1(L^{-2})
}
computes what is called the Zarisski tangent cone. The dimension of this
space is $4k-4$ --- i.~e.~it is bigger by two than the dimension of the
moduli space.
The fields corresponding to the cohomology groups \zaris\ are
charged with
respect to $U(1)$: the elements of $H^1(L^2)$ have charge $+2$ and the
elements of $H^1(L^{-2})$ have charge $-2$. The matter should fit into
$N=2$ hypermultiplets, so it is necessary that ${\rm dim}\,H^1(L^2)={\rm
dim}\,H^1(L^{-2})$. This condition is indeed satisfied due to Serre
duality on $K3$. Thus the $U(1)$ theory has $2k-2$ hypermultiplets with
charges $\pm 2$. On the Coulomb branch away from the Higgs branch these
hypermultiplets are massive. The dimension of the $U(1)$ Higgs branch is
$2(2k-2)-2=4k-6.$ It coincides with the dimension of the moduli space of
$SU(2)$ instantons, as necessary for consistency.

We conclude that the moduli space of $SU(2)$ bundles on $K3$ with  even
instanton number $k$ can be described as a baryonic Higgs branch of
$SU(2)$ theory with
$k$ flavors in the vicinity of  pointlike instantons and as a Higgs
branch of $U(1)$ theory with $2k-2$ charged hypermultiplets in the
vicinity of reducible bundles $L\oplus L^{-1},\ c_1^2(L)=-k$.

The situation  for $SU(N),\ N>2$ bundles is the straightforward
generalization
of the construction discussed above. In the vicinity of the
reducible bundles
$V_n=\oplus_i V_{r_i}$ the $SU(n)$ baryonic Higgs branch can also
be described
as  Higgs branch of  a different gauge theory with the gauge group being
$\otimes SU(r_i)$ times appropriate number of $U(1)$s.

\newsec{$K3 \times K3$ vs $CY_3 \times T^2$ compactifications}
\subsec{A Type IIA dual}

Let us start with  a (rather degenerate) situation in
which the
heterotic bundle on $T^2\times K3$ is a
 pullback of the bundle $V_1\times V_2$ on $K3$. The structure
group of this
bundle is  $H_1 \times H_2$, we assume that $H_i$ are both simple $ADE$
groups. The second Chern class is distributed between $V_1$ and $V_2$ as
$(12+n, 12-n)$.
The unbroken gauge group in four dimensions  is
$G_1 \times G_2$ (where $H_i \times G_i \subset E_8$ is a maximal
embedding)\foot{To be precise, this statement is true only classically.
The relative locations of the 7-branes parameterize the {\it
Coulomb} branch
that gets quantum correction.}. Each spectral surface $\S(V_i)$ has
a  multiple
component equipped
with the $H_i$-bundle $V_i$.

The F-theory dual is
compactified on $K3 \times K3$, where the first factor is a {\it
singular}
elliptic $K3$. The elliptic fibration has two $E_8$ singularities and
generically four $I_1$ singularities.  The two  7-branes with $E_8$
singularities are equipped with the $H_i$-bundles $V_i$. The $E_8
\times E_8$
gauge group is
broken by
the $H_1 \times H_2$ instantons leaving the unbroken gauge group
$G_1 \times G_2$ times the appropriate number of $U(1)$s.
The moduli space of the gauge fields inside the 7-branes and the gauge
fields inside the spectral surface are isomorphic to each other.

The $E_8$ singularity in F-theory can be deformed away, but the smallest
possible singularity is fixed by the group $H_i$. The $H_i$ 7-brane
 is equipped
with the irreducible $H_i$-bundle $V_i$, so it cannot split. The
deformations
destroying the $E_8$ singularities describe the {\it classical}
Coulomb branch
of the $N=2$ susy theory with  the gauge group $G_1\times G_2$. To
obtain the
quantum Coulomb branch, one needs to account for the
nonperturbative effects.

For example, consider two 7-branes
located close to each other on \Pt1\ and wrapped on $K3$. Separating
the center
of mass, we get a Coulomb phase of  the $SU(2)$ gauge theory with no
matter in
four dimensions\foot{Assuming that the 3-branes are located far away
from these
7-branes so that the corresponding hypermultiplets are massive.}.
Quantum mechanically, the mass gap $\Lambda^2=e^{-{V/ g^2_{\rm
str}}}$ is
dynamically
generated, and the
coupling constant is given by the Witten-Seiberg formula
\eqn\coco{
{1\over g^2}={V\over g_{\rm str}^2} + {\rm log}(a^2)+
\sum_{N=1}^{\infty}c_N e^{-2N({V\over g_{\rm str}^2} + {\rm log}
(a^2)) },
}
where $V$ is a volume of $K3$ and $a$ is a scalar of the $U(1)$ vector
multiplet. The first term in \coco\ gives the perturbative
effective tension of
a pair of  7-branes wrapped around $K3$ and separated by $a$ on
\Pt1. The ${\rm
log}\,a^2$ term is a contribution of the (Euclidean) gas of massive
strings
stretched between the 7-branes.
The exponential terms in \coco\ come from the Euclidean 3-branes
wrapped around
$K3$. When the 3-brane is wrapped on top of the 7-brane it gets
additional
zero modes coming from the massless string stretched between the
3-brane and
the 7-brane, which make it contribute to the prepotential.

In the semiclassical regime $\mid a\mid^2 >> \Lambda^2$
the scalar
$a$ measures the separation between the 7-branes. However, in
quantum theory
there is no point
on the
Coulomb branch where $a=0$  and the
$SU(2)$ is never
restored \ref\seiwit{N. Seiberg and E. Witten, Nucl.Phys. {\bf
B426} (1994)
19}.

This F-theory compactification has  another interpretation. The
$(12-n, 12+n)$
heterotic compactification on $K3$ is
equivalent to the F-theory
compactification on the Calabi-Yau threefold $CY_3$. The threefold
$CY_3$ is an
elliptic fibration over the Hirzerbruch
surface
${\bf F}_n$. Compactifying further on $T^2$ we get an   F-theory
compactification on
$T^2\times CY_3$ which can also be interpreted as a Type IIA
compactification
on $CY_3$. Either one  has to be dual to the above
  $K3 \times K3$ compactification.

As a result, we arrive to the conclusion that type IIB on
$P^1 \times K3$ with 7-branes wrapped on $K3$ is equivalent to the
type IIB on
${\bf F}_n  \times T^2$  with 7-branes wrapped on $C_i\times T^2$, where
$C_i$ are
curves in ${\bf F}_n$!  Geometrically, these two compactifications seem to
be very
different. For example, one manifold is simply connected while the
other one
has non-contractible paths.  It is
remarkable
that one can trade the
topology of the manifold for the non-trivial gauge bundles on 7-branes.

It is
instructive
to compare the
 low energy descriptions. Both theories are $N=2$ super YM
with the
gauge
group $G_1\times G_2$ coming from 7-branes. The ${\bf F}_n\times T^2$
compactification has a $G_1$ 7-brane wrapped on the zero section of
${\bf F}_n$ and a
$G_2$ 7-brane wrapped at the section at infinity. These 7-branes are also
wrapped on $T^2$. In this language, the Coulomb branch is
parameterized by the
$G_1\times G_2$ Wilson lines on the torus. In the Type IIA language, the
Coulomb branch corresponds to  the K\"ahler classes of the components of
exceptional fibers over the discriminant locus with $G_i$ singularity.
Comparing this with the description in terms of Type IIB on
$\P1\times K3$ we
see that the {\it complex deformations} of the elliptic fibration
over \Pt1\
are mapped to the {\it K\"ahler moduli} of $CY_3$. The quantum
K\"ahler moduli
space is corrected by the worldsheet instantons. Translating this back to
F-theory we get the ``quantum elliptic fibration'' which describes
the same
space in terms of complex deformations of something.

The description of the matter multiplets also  differs between two
languages.
On the one hand, the ${\bf F}_n\times T^2$  compactification gets the
hypermultiplets
from the 7-branes intersecting along tori $T^2$
\ref\bsv{M. Bershadsky, V. Sadov and C. Vafa, Nucl. Phys. {\bf
B463} (1996)
398}. The Higgs branches
corresponding to the nonzero $vev$'s of these fields describe the complex
deformations of $CY_3$ destroying the $G_1\times G_2$ singularity.
On the other
hand, in the $\P1\times K3$ compactification the matter
hypermultiplets come by dimensional reduction from 8 dimensions.
The number of matter multiplets $N_i$ in the representation $R_i$
of the unbroken group $G$ is given by the index theorem
$N_i=(12 \pm n) {\rm index}(S_i)-{\rm dim}(S_i)$, where $S_i$ is the
representation of $H$ entering into the decomposition ${\bf 248}=\oplus_i
(R_i,S_i)$. This computation is identical to the computation in the
heterotic string. The Higgs branch can be described completely in
terms of
the bundles $V_1$ and $V_2$. Therefore,  the moduli of $V_1$ and
$V_2$ are
mapped into the {\it the complex moduli} of $CY_3$.

\subsec{3-branes vs blowups}

The theories labeled by different values of $n$ are related to each
other by
phase transitions. From the heterotic theory point of view one has
to shrink
one instanton (say of $V_1$) to zero size, reinterpret it as a
5-brane and then
``dissolve'' it into a  finite  instanton of the second bundle.
Similarly,
in the description of F-theory compactified  on $K3 \times K3$  the
process of
changing $n$ consists of three steps.
 First, one shrinks  the instanton inside the first 7-brane down
to zero size and replaces it by a 3-brane. The 3-brane can move
freely inside
$\P1 \times
K3$. Then one
 puts a 3-brane on top of another 7-brane and finally dissolves it into a
finite size
instanton.

If one removes sufficiently many instantons from one of the 7-branes,
so that the
instanton number $k$ is less then $10$, the background bundle does not
break $E_8$
completely and the enhanced gauge symmetry appears.  For $k=3$, the group
is $SU(3)$, for $k=4$, it is $SO(8)$ etc.

In
the F-theory on $T^2\times CY_3$ (or equivalently, in Type IIA on $CY_3$)
 this transition can be described as a sequence of one blowup and
one blowdown
of the base ${\bf F}_n$.
Let us blow up a point on the zero
section of ${\bf F}_n$. Two rational curves pass through this
point: one is the
zero section
$S_0$ and the other is the fiber $F=\P1$ of the projective bundle
${\bf F}_n$. These
curves generate the cohomology ring of ${\bf F}_n$. They satisfy $F^2=0,\
 S_0^2=n,\
S_0\cdot F=1$. After the blowup $\pi : B\rightarrow {\bf F}_n$, the
cohomology ring
of $B$ is generated by $\pi^*F, \pi^*S_0$ and the exceptional
divisor $E$ such
that $E^2=-1,\ E\cdot S_0=E\cdot F=0$. Two curves $S_0$ and $F$
which pass
through the blown up point are transformed into
$\hat{S_0}=\pi^*S_0-E$ and
$\hat{F}=\pi^*F-E$.
Their intersections are: $\hat{S}_0^2=n-1,\ \hat{F}^2=-1,\ \hat{F}\cdot
\hat{S_0}=0$ and $\hat{F}\cdot F=0$. These relations show that the line
$\hat{F}$ can be blown down which results in $B\rightarrow
F_{n-1}$. The new
zero section is $\hat{S_0}$.
This process
changes the index
of the Hirzebruch surface $n \rightarrow n-1$ (or $n \rightarrow
n+1$, if we start with a point on $S_\infty$).

Blowing up a point on the curve $S$ with self-intersection $S^2=-k$ gets
this curve properly transformed into a curve $\hat{S}=\pi^*S-E$ with
self-intersection $-(k+1)$. If this number is less than $-2$, the curve
$\hat{S}$ has to be a component of the discriminant locus \mv, as a
consequence of the adjunction formula which tells us that the  
intersection of
the rational curve $S$ with the discriminant $\Delta =-12K$ is given by
$\Delta \cdot S=12(S^2+2)$. For $S^2 < -2$, $\Delta \cdot S<0$ and since
$\Delta$ is an effective divisor the curve $S$ has to be one of its  
components:
$\Delta =\Delta' +rS$, where the integer $r$ satisfies

\eqn\ineqr{
r\geq 12{S^2+2\over S^2}
}

The type of the singular elliptic
fiber over $S$ is also determined by $S^2$.
For example, the self-intersection $-3$ corresponds to the $I_3$
fiber and
the unhiggsed $SU(3)$ gauge group in space-time and the self-intersection
$-4$ corresponds to the $I^*_0$ fiber and the unhiggsed $SO(8)$.  
The gauge
group $E_6$
appears
for self-intersections $-5,\, -6$, the group $E_7$ -- for  
self-intersections
$-7,\, -8$, and
the group $E_8$ -- for self-intersections $-9,\, -10,\, -11,\,  
-12$.In general,
the charged matter that could higgs the gauge group in space-time  
comes from
the intersection $\Delta'\cdot S$ of $S$ with the other components of
discriminant. So whenever $\Delta'\cdot S=0$, the gauge symmetry cannot be
higgsed. This condition implies that the inequality \ineqr\ is  
saturated which
can only happen
for self-intersections $-3,\, -4,\, -6,\, -8,\, -12$. In the language of
 the Calabi-Yau elliptic fibration $CY_3$ this means that there are no
deformations
destroying such a component of the discriminant locus, so  
everywhere in the
moduli space of
$CY_3$
the base of the elliptic fibration ought to have a curve with
self-intersection $-3$, $-4$ etc.
We will discuss the consequences of that in the next section. The minimal
gauge group which cannot be Higgsed is the same as the minimal  subgroup
of $E_8$, left unbroken by $12+S^2$ instantons.

Comparing two pictures, we conclude that the F-theory on $K3\times K3$
with $N=m+n$ 3-branes and two $E_8$ 7-branes with the instanton numbers
$12-m$ and $12-n$  is described by the Type IIA compactified on $CY_3$
elliptically fibered over the base which is $\P1 \times \P1$ blown up in
$m+n$ points. The set of $m$ points belongs to the line $S_0$ and the set
of $n$ points belongs to the line $S_\infty$ which does not intersect
$S_0$. The blown up surface has two nonintersecting rational curves with
with self-intersection $-m$ and  $-n$ respectively.

The  vector multiplets in Type IIA which come from the
exceptional divisors
$E_i$ correspond to the vector multiplets on  3-branes. The
complexified area of
$E$ should be interpreted as the \Pt1\ coordinate of the 3-brane on
$\P1\times K3$.   Let us denote by $\omega$
the K\"ahler class of the base. Then
\eqn\cons{\omega(E)+\omega(\hat F)=\omega(F)={\rm const}~.}
The variation of $\omega(E)$ from zero to $\omega(F)$ corresponds to the
3-brane
motion from one $E_8$ 7-brane to another.

The $E_8$ hypermultiplet corresponding in the F-theory language to
a string
connecting a 3-brane with the first $E_8$ 7-brane, in the Type IIA
language
should correspond to 2-branes wrapped around the rational curves on the
exceptional divisor $D_1\subset CY_3$ which covers the exceptional curve
$E\subset {\bf F}_n$. Similarly, the hypermultiplet corresponding
to the string
connecting the 3-brane with the second $E_8$ 7-brane corresponds to
the Type
IIA 2-branes wrapped around the divisor $D_2$ covering the exceptional
curve $\hat{F}$. Both $D_1$
and $D_2$ are
 almost Del Pezzo surfaces with $\chi(D_i)=12$, so that the
supersymmetric 2-cycles in each of them
reproduce the $E_8$
lattice.

It should be emphasized  that since the 3-brane description
actually gives the
{\it quantum} Coulomb branch, the above map is  actually a
{\it mirror map}. It translates the worldsheet instanton corrections in
Type IIA into the nonperturbative prepotential in the effective $N=2$
four-dimensional theory on the 3-brane probe.

\subsec{Del Pezzo surfaces and elliptic fibrations}

Now consider the Type IIA compactification on a Calabi-Yau
threefold $CY_3$
which is an elliptic fibration over the base $B$, where $B$ is
obtained by
blowing up $N\leq 24$ points on ${\bf F}_n$. This compactification is
dual to the
F-theory compactification on $K3\times K3$ with $N$ 3-branes in the bulk.
In the previous section we discussed the map between the positions
of 3-branes
on \Pt1\ and the (complexified) blowup  K\"ahler moduli of $CY_3$.
The other
K\"ahler moduli of $CY_3$ come from the exceptional fibers of the
elliptic
fibration. They correspond to the Coulomb branch of the $N=2$ gauge
theory on  7-branes.

Let us start by blowing up $1$ point $P$ on $F_1$. Depending on
whether or not
this point lies on the line $S_\infty$, the result will be
different. If $P$
does not lie on $S_\infty$, the blown up surface $B_2$ will have
two $(-1)$
curves: $S_\infty$ and the exceptional divisor $E$. Alternatively,
if  $P\in
S_\infty$, the $(-1)$ curve $S_\infty$ gets properly transformed
into a $(-2)$
curve $C=S_\infty-E$, so the blowup $\tilde{B}_2$ has one $(-1)$
curve $E$ and
one $-2$ curve $C$. The surfaces $B_2$ and $\tilde{B}_2$ have
different complex
structures. The $(-2)$ curve $C$ in $\tilde{B}_2$ corresponds to the
root of
$SU(2)$. If $C$ is contracted down to zero size, a nodal $A_1$
singularity
forms on $\tilde{B}_2$, and the Calabi-Yau 3-fold
elliptically fibered over $\tilde{B}_2$ gets an elliptic curve of $A_1$
singularities.

Blowing up 2 points $P_1$, $P_2$ on $F_1$, we have more
possibilities. We are
mostly interested in $(-2)$ curves, so the following cases are worth
mentioning.
\item{i)} Both $P_1\in S_\infty$ and $P_2\in S_\infty$, $P_1\neq
P_2$. Then the
curve $S_\infty$ is transformed into a $(-3)$ curve
$\S=S_\infty-E_1-E_2$.
There are no $(-2)$ curves on the blowup. The curve $\S$ becomes a
component of
the discriminant locus of $CY_3$ with the singularity $A_2$, so
there is $N=2$,
$SU(3)$ Yang-Mills theory in space-time.
\item{ii)} $P_1\in S_\infty$, $P_2\notin S_\infty$ and $P_1, P_2$
do not belong
to the same fiber $F$. Then the only $(-2)$ curve on the blowup is
$C=S_\infty
- E_1$.
\item{iii)} $P_1,P_2 \notin S_\infty$ and both $P_1$ and $P_2$ lie
on the same
fiber $F$. Then $F$ is properly transformed into a $(-2)$ curve
$F-E_1-E_2$.
\item{iv)} $P_1=P_2\notin S_\infty$. Considering one blowup after
another, we
end up either with {\it one} $(-2)$ curve $E_1-E_2$ if $P_2$
approaches $P_1$
generically  or with {\it two} nonintersecting  $(-2)$ curves
$E_1-E_2$ and
$F-E_1-E_2$ if $P_2$ approaches $P_1$ along the fiber $F$.
\item{v)} $P_1=P_2\in S_\infty$.
\itemitem{a)} If $P_2$ approaches $P_1$ along
$S_\infty$,
there is only one $(-2)$ curve $E_1-E_2$, and a $(-3)$ curve $S_\infty
-E_1-E_2$ which has $A_2$ singular elliptic fiber over it.
\itemitem{b)} If $P_2$
approaches
$P_2$ along generic direction, there are two $(-2)$ curves
$S_\infty-E_1$ and
$E_1-E_2$ forming the root system of $SU(3)$.
\itemitem{c)} Finally, if $P_2$
approaches
$P_1$ along the fiber $F$, one gets the third $(-2)$ curve
$F-E_1-E_2$ and the
root system of $SU(2)\times SU(3)$.

All  examples  (i-v) correspond to two instantons  shrinking to  
zero size. In
this interpretation all these cases differ by the relevant  
orientation of the
instantons inside the $E_8$. For example, in the cases (i), (v.b)  
the $SU(3)$
subgroup is restored, leaving $E_6$ completely  broken by
the remaining instantons.

The number of possibilities increases with the number of blowups.
For example, for 3 blowups the  $A_4$ root system (maximal )
appears when one
makes 3 blowups on $S_\infty$ and top of each other. Concretely, $P_1\in
S_\infty$, $P_2$ approaches $P_1$ along the fiber and $P_3$ approaches
$P_2$ from  generic direction. The $A_4$ root system is
generated by
four $(-2)$ curves $S_{\infty}-E_1,~E_1-E_2,~E_2-E_3$ and
$F-E_1-E_2$.  In
general, for $N \leq 8$ blowups the maximal root
system coincides with the root system of $E_N$ Lie algebra, where
$E_3=A_2 \times A_1, ~E_4=A_4$ and $E_5=D_5$ (see the discussion in
\ref\doug{Michael Douglas, Sheldon Katz and Cumrun Vafa, {\it Small
Instantons,
del Pezzo Surfaces and Type I' theory}, hep-th/9609071}).

Already these simple examples allow us to learn some lessons. Let
us test further the correspondence between 3-branes on $\P1\times K3$ and
blow-ups on $F_1$. When  $k$ 3-branes meet, we expect to see $N=4$,
$SU(k)$ supersymmetric Yang-Mills theory in space-time. Also, if the
number of instantons on the $E_8$ 7-brane is less than $10$, a part of the
full $E_8$ group should be restored. In Type IIA picture this appears to
correspond to the situation when one blows up $k$ points $P_i\in
S_\infty$
(for $k=2$, this is the case v.a above). To get $SU(k)$, one needs to
bring the points $P_i$ together, moving them along $S_\infty$. The root
system of $SU(k)$ is generated by the $-2$ curves $E_i-E_{i+1}$.  
Shrinking
this system of curves to a point one produces an $A_{k-1}$ singularity at
that point on $B_{k+1}$ and a whole elliptic fiber of $A_{k-1}$
singularities in the elliptic fibration $CY_3$.  In Type IIA
compactification such singularity is known to correspond to $N=4$,
$SU(k)$
supersymmetric Yang Mills.
Also, the line $S_\infty$ with $k$ points blown up becomes a $(-k-1)$
curve $S_\infty - \sum_i E_i$. For $k\geq 2$, it has to be a component of
the discriminant with the singular fiber which correctly describes the
subgroup of $E_8$ left unbroken by $11-k$ instantons.

It should be noted that the complex structure of the blow-up $B_{k+1}$
depends on the relative positions of $P_i$. For $k\leq 3$, the blow-ups
are rigid\foot{This follows from the fact that there are three different
${\bf P}^2$ with four marked points, distinguished by how many points (2,
3 or 4) lie on one line.} and one has a discrete set of different
$B_{k+1}$'s.
 For $k\geq 4$, the generic surface $B_{k+1}$ has $2(k-3)$ complex
deformations. The generic surface has no $(-2)$ curves which
appear along
certain divisors in the moduli space. The moduli space of the base should
be considered as  a part of the moduli space of the Calabi-Yau fibration
$CY_3$. It turns out that moving along this moduli space, one cannot
connect the generic surface $B_{k+1}$ with some blowups. The most
important example is the blowup of $F_1$ in $>1$ points  on $S_\infty$ or
$>3$ points on  $S_0$, so that there is a curve with self-intersection
less than $-2$. As we explained in section 5.2, $CY_3$ has no  
deformations
destroying a $-3$ curve in the
base, so we can never reach a generic surface $B_m$ which has no  
such curves.
Similarly,
$-4$, $-6$, $-8$ and $-12$ curves cannot be destroyed. Thus the
theory where
 $N$ generic points on the base are blown up  cannot be
reached starting from the theory where the points which are blown
up lie on two
curves.

Blowing up 24 points on two lines in $\P1 \times \P1$ lands us on the
$(43,43)$ Calabi-Yau fibration, which has two rational curves of generic
$E_8$ singularities \vws, \ref\can{P. Candelas, A. Font, {\it  
Duality Between
Webs of Heterotic and Type II Vacua}, hep-th/9603170}.
This compactification is equivalent to F-theory on
$K3 \times K3$ with the first $K3$ having two $E_8$ singularities and 24
3-branes in the bulk.

On the other hand, one can blow up as many as $8$ {\it generic} points on
$F_1$ and that would land us on the $(19,19)$ Calabi-Yau.
This threefold is a double elliptic fibration over
\Pt1, so that
the fiber is a product of two
elliptic curves. The K\"ahler cone of $CY_3$ is generated by the  
K\"ahler cones
of
$B$ and $B'$,
with a single relation coming from the class of \Pt1\ shared by
both $B$ and
$\tilde B$. It is clear that this compactification is connected through a
series of phase transitions with the model $(43,43)$. Indeed, one  
can blow down
all $24$ spheres in $(43,43)$ model, get type IIA compactification on
Calabi-Yau threefold fibered over ${\bf F}_1$ and then blow up eight  
points on
${\bf F}_1$.
It follows  that one should be able to see the $(19,19)$  
phase in the
F-theory description. We conjecture that some degenerations of  
gauge bundle
inside the 7-branes lead to this phase.
The relevant orientation of the instantons shrinking to zero size  
is crucial: the instantons should be embedded in such a way that by  
shrinking each
of them to zero size we adjust every time exactly $29$ parameters.  
These instantons should break $E_8$ to  
an  abelian
subgroup.

\newsec{Acknowledgments}

We are grateful to  Andrei Johansen, Tony Pantev and Cumrun Vafa
for many useful discussions. The research of M.~B.~ was partially
supported by
the
NSF grant PHY-92-18167, the NSF 1994 NYI award and the  DOE 1994 OJI
award. The research of V.~S.~was supported in part by the NSF grants DMS
93-04580, PHY 9245317 and by Harmon Duncombe Foundation.

\listrefs

\end